\documentclass[acmlarge]{acmart}

\usepackage{subfiles}

\AtBeginDocument{%
  }

\setcopyright{acmlicensed}
\copyrightyear{2026}
\acmYear{2026}
\acmDOI{XXXXXXX.XXXXXXX}
\acmConference[FAccT '26]{The 2026 ACM Conference on Fairness, Accountability, and Transparency}{June 25--28, 2026}{Montreal, QC, Canada}
\acmBooktitle{The 2026 ACM Conference on Fairness, Accountability, and Transparency (FAccT '26), June 25--28, 2026, Montreal, QC, Canada}

\begin{document}

\title{Can Data Work be Reparative?}

\author{Srravya Chandhiramowuli}
\email{srravya.c@ed.ac.uk}
\affiliation{%
  \institution{University of Edinburgh}
  \city{Edinburgh}
  \country{UK}
}

\author{Ding Wang}
\email{drdw@google.com}
\affiliation{%
 \institution{Google Research}
 \city{Atlanta}
 \country{USA}
 }

\author{Alex S. Taylor}
\email{alex.taylor@ed.ac.uk}
\affiliation{%
  \institution{University of Edinburgh}
  \city{Edinburgh}
  \country{UK}
}

\renewcommand{\shortauthors}{Chandhiramowuli et al.}

\begin{abstract}
We present an ethnographic study of an alternative approach to data work, developed by a civic-tech initiative that builds datasets for training and benchmarking online safety systems. They aim to respond to online safety concerns from a feminist perspective, by building safety datasets collaboratively with those most impacted by online harms. In this paper, we examine how this approach aims to reorient data work as a site for repair and redress, and trace the struggles they encounter in the process. Specifically, we draw attention to the challenges and tensions involved in advancing just reward for data work and collective governance of AI datasets. Examining these challenges through an STS-informed lens of reparative justice and repair, we argue that the work of repairing data work (and AI) lies, fundamentally, in resetting the ties of accountability. At a time heightened emphasis on efforts like safety evaluations and red teaming to make AI more “responsible”, we highlight the need to confront foundational questions about how the humans involved in these efforts relate to the datasets and systems they help produce. A reparative lens demands that we interrupt prevailing norms of data work and place at their centre, not AI or datasets, but those most harmed by the neglect, oversight and exclusion animated in the current modes of dataset production. This, we argue, offers a bold vision for responsibility and contributes towards a critical agenda for building alternative futures of data and AI practice. 
\end{abstract}




\maketitle

\section{Introduction}

Even as contemporary computing systems register significant advances in their computational capabilities, contextual human input provided through a range of data work tasks and activities is indispensable for training, fine-tuning, and evaluating these systems \cite{gray_ghost_2019, roberts_behind_2019, tubaro_trainer_2020}. Yet, the dominant modes of data and AI production require a data work that is routine and repetitive \cite{chandhiramowuli_making_2024, zhang_making_2025, miceli_data-production_2022} producing what Sambasivan et al \cite{sambasivan_everyone_2021} refer to as cascading issues in datasets. Moreover, data workers, treated as a homogeneous collection of interchangeable actors \cite{diaz_crowdworksheets_2022}, are too often assigned severely limited roles with little scope for applying individual expertise and skill. The growing recognition of the restrictive structures and difficult conditions \cite{bennett_everybody_2025, catanzariti_taming_2025, miceli_between_2020, posada_deeply_2024, rothschild_problems_2024, wang_whose_2022, yang_guilds_2024} underscores the need to interrupt entrenched patterns of labour exploitation, knowledge extraction, and epistemic imposition in dataset production. This, in turn, warrants attending to alternative approaches to data work that seek to repair and reorient data work towards a more just and equitable practice. 

In this paper, we examine the possibilities for such change through a case study of a civic-tech initiative that builds datasets for training and benchmarking online safety systems. We present the case study of Tattle Civic Tech, an organisation based in India, building machine learning datasets and tools to counter online harms such as misinformation, hate, and online gender-based violence (oGBV). They aim to respond to online safety concerns from a feminist perspective, by building safety datasets collaboratively with those most impacted by online harms. That is, rather than seeking to eliminate human subjectivity, this approach actively incorporates it in datasets. We examine this alternative orientation to dataset production, specifically focusing on the challenges and tensions they encounter in addressing the uneven structures and relations of power in data work. By foregrounding these challenges, we identify the open questions and avenues for intervention that are crucial to confront and consider critically in developing alternative approaches to data work. 

Drawing on our ethnographic research at Tattle, we elaborate on two dataset projects that address content moderation and AI safety needs in Indian languages. The two datasets---a lexicon of gender abusive slurs in four Indian languages and an LLM safety benchmark dataset in Hindi---contribute much needed, yet sorely lacking, resources on the specific linguistic, regional and cultural nuances of harms in non-English languages. Tattle strives to adopt a feminist approach to this work: the two datasets on online harms were produced collaboratively through invitations to contributors who have lived experiences encountering or responding to the harms. Those involved so far have included social workers, activists, researchers, journalists, and psychologists directly engaged with the harms, and living and working in the specific regional, linguistic and cultural contexts in which harms are experienced. Building on their intimate familiarity with the harms and a commitment to address them, they contribute directly to the datasets by adding and annotating data on the harms. In other words, the contributors perform data work. 

In developing this approach to safety datasets, Tattle seeks to reimagine content moderation and online safety—as a collaborative endeavour grounded in carefully curated forms of knowledge, and expertise. This necessarily includes rethinking how data is collected, labelled, generated, and categorised in the datasets. Far from being limited within the narrow confines of tasks and labels, their contributors shape how harms and safety are conceptualised and rendered in the datasets. To facilitate this deeper engagement, Tattle structures dataset contribution activities in ways that lower barriers to participation for contributors, nurture discussions and collaboration among them, and are attentive to the toll of working with toxic content. While their approach demonstrates possibilities for recovering data work from reductive frames, it also opens up a new set of questions and challenges for reorienting data work, that we elaborate in this paper.

To examine the open challenges surfaced in the case study, and the productive avenues they bring into focus for repairing and reorienting data work, we draw on scholarly work surrounding reparations and reparative justice \cite{benjamin_captivating_2019, browne_dark_2015, walker_moral_2006} as an analytical frame. Through this scholarship, reparations combine ideas of repair and justice to advance a distinct orientation that centres the lives and experiences of the oppressed, persecuted and stigmatized \cite{walker_making_2015,walker_what_2010}. This has found resonance in Science and Technology Studies (STS), with scholars developing conceptual links to reparative justice in the theorising of algorithmic redress \cite{davis_algorithmic_2021} and ecological care and repair \cite{papadopoulos_ecological_2023}. Building on this line of inquiry, we ask what we see to be foundational questions for justice-orientated, reparative data work: can data work be a mode of repair and redress for injustice encountered in online harms? How, if at all, might data work be reparative and what might a reparative data work seek to repair?  

Tracing Tattle’s dataset projects ethnographically, our findings highlight the challenges confronted by Tattle in attempting to craft an expansive, collaborative role for their dataset contributors. First, recognising the contributors as experts and their input as crucial and valuable to the datasets requires compensating them as such. Here, we draw attention to the tensions in moving beyond minimum standards (such as minimum wage and living wage) and identifying what constitutes just reward for data work. Further, rethinking recognition for dataset contributors invites questions about their relationship to the datasets they help produce. This includes considering whose views shape dataset decisions, what rights contributors have over their contributions and what the role of Tattle might be as custodians of the datasets. By elaborating on Tattle’s exploration for suitable models of collective dataset governance to confront these questions, we highlight the open challenges and tensions in bringing alternative visions for AI into actual practice. 

Amidst calls for reimagining AI and technofutures, it is pertinent to attend to the work that lies ahead in operationalising alternative visions. This research we present in this paper takes a step in this direction, offering empirical insights into an alternative approach in the making that open up a crucial line of inquiry (rather than seeking to resolve it) for “do[ing] data work otherwise” \cite{disalvo_when_2024}. Examining these empirical insights through the analytical lens of reparative justice, we argue that the work of repairing data work lies, fundamentally, in resetting the relations of accountability in AI. We emphasise the need to define shared standards and forms of responsibility that interrupt prevailing norms of dataset production and place at their centre, not AI or datasets, but those most harmed by them. At a time when much emphasis and indeed, investments are placed on efforts like safety evaluations and red teaming to make AI more “responsible”, we highlight the need to confront foundational questions about how the humans involved in these efforts relate to the datasets and systems they help produce. We reflect on these broader implications in the discussion section.

Our paper makes three main contributions to critical scholarship concerned with data work, and ethics and responsibility in AI: 
\begin{enumerate}
    \item We contribute valuable empirical insights into the challenges and tensions entailed in developing an alternative approach to data work, particularly in/from the global south. In doing so, we extend Responsible AI and AI ethics scholarship beyond its predominant focus on Western, Educated, Industrialised, Rich, and Democratic (WEIRD) contexts \cite{septiandri2023weird}.
    \item We provide insight into how labour concerns in data work are entangled with the politics of knowledge production. We demonstrate, through the study of Tattle, how concerns of epistemic justice and labour justice are deeply entangled and why they cannot (and should not) be viewed in isolation. This bolsters calls for AI/ML researchers and practitioners to recognise concerns that go beyond, yet interact with labour conditions \cite{miceli_between_2020, wang_whose_2022}
    \item Finally, our paper demonstrates the analytical value in thinking with reparative justice to inform efforts to reorient data work and AI production. Reparations demand an unwavering commitment to redress and justice for those most harmed by systemic injustice and violence; this offers a bold vision for responsibility, one that might serve as an anchor for ongoing endeavours in responsible AI.
\end{enumerate}

\section{Related Work}

\subsection{Alternative visions of AI}

Tattle’s work can be positioned within a broader set of initiatives that envision alternative approaches to AI. They stem from a growing recognition of the limits and issues with dominant modes of AI production that result in bias, discrimination, and exclusion, as well as exploitation, extraction and concentration of wealth. This has motivated calls for rethinking and repairing how data and AI-based technologies are envisioned, developed and put to use. Perhaps the loudest of those calls, and certainly one whose language has been widely adopted across the tech industry, public sector and research, is ‘Responsible AI’ (R-AI). As Tollon and Vallor \cite{tollon_responsible_2025} note, Responsible AI does not refer to one particular thing but encompasses a set of interweaving goals and approaches that includes fairness, transparency, accountability, ethics, safety, privacy and trustworthiness. While R-AI goals received enthusiastic support and commitment of resources initially, the support and commitments have since been in retreat, in favour of AI market growth and geopolitical dominance \cite[p.~11]{tollon_responsible_2025}. 

A second strand of efforts to reimagine AI are in the domain of participatory AI. It advocates bottom-up approaches to de-centralise decision-making and prioritise the involvement of those most adversely impacted by AI \cite[among others]{birhane_power_2022, delgado_participatory_2023, sieber_what_2024, suresh_participation_2024, young_participatory_2025}. In recent years, participatory methods have steadily gained recognition as a valuable means to better understand and account for experiences that tend to be marginalised in AI development and governance \cite{coffey_maori_2021, hao_new_2022, hu_enrolling_nodate, katell_toward_2020, lee_webuildai_2019, nekoto_participatory_2020, suresh_towards_2022, queerinai_queer_2023}. While a range of initiatives demonstrate possibilities for widening consultation around particular design choices as well as co-creating them, scholars note the limits of participatory AI in ceding power \cite{delgado_participatory_2023, prabhakar_participation_2023, sloane_controversies_2024, young_participation_2024} and warn against risks of dilution and co-option of participatory efforts \cite{birhane_power_2022, sloane_participation_2022, suresh_participation_2024}. This line of critique highlights the power imbalances within which efforts to reorient AI must operate, even as they seek to shift that imbalance. To contend with this tension and develop a generative analysis for how we might address it, we turn to lessons from reparations and reparative justice. 

\subsection{Thinking with reparative justice}

Reparations are typically associated with compensation provided to victims of human rights violations such as during war and conflict, or historical injustice such as colonisation and slavery. They signify a commitment to acknowledge and repair past injustice, and affirm renewed relationships of trust, respect and reconciliation \cite{walker_making_2015, walker_what_2010} for the future. In practice, this translates into measures providing recognition, concrete relief, and direct support to those who have suffered injustice. In analytical terms, reparations combine ideas of repair and justice to advance a distinct orientation that centres the lives and experiences of the oppressed, persecuted and stigmatized \cite{walker_making_2015}. 

This has found resonance in STS, with scholars developing conceptual links to reparative justice in theorisations of algorithmic redress \cite{davis_algorithmic_2021} and ecological resurgence \cite{papadopoulos_ecological_2023}. Building on this line of scholarship that demonstrates the conceptual value of reparations, we draw on reparative justice as an analytical frame to examine efforts that seek to interrupt and reimagine dataset production and AI. In particular, we think with feminist philosopher Margaret Walker’s theorising \cite{walker_making_2015, walker_moral_2006, walker_transformative_2016, walker_what_2010}, in which she emphasizes that the moral and political commitments of reparative justice lie, not in undoing past harms, but in addressing the threat of neglect, erasure or contempt faced by those harmed. To repair this, she argues, those responsible must \textit{"acknowledge the reality and the nature of the wrongs done, and appreciation of the insult and harm suffered, by the victim in wrongful treatment; they must affirm the victim’s deservingness of repair as a matter of justice and the attempt at repair as an obligation of justice.”} \cite[p.~20]{walker_what_2010}. The goals of reparative justice lie in demonstrating---in concrete, actionable terms---a shared understanding of accountability and just relations in the present that can instill hope for a better, transformed future.

Recent scholarship in critical HCI, STS and allied fields has demonstrated the relevance of repair and reparative justice for technological and algorithmic contexts \cite{eglash_computational_2024, houston_values_2016, papadopoulos_ordinary_2023, lin_bias_2023, peterson-salahuddin_repairing_2024, rakova_terms-we-serve-_2023, tracey_after_2024}, building conceptual linkages between care \cite{de_la_bellacasa_matters_2011}, repair \cite{jackson_rethinking_2014} and notions of redress and justice \cite{davis_algorithmic_2021, papadopoulos_no_2023}. Davis et al \cite{davis_algorithmic_2021} propose algorithmic reparations as a framework for critical algorithmic reform that prioritises redress over idealism. Their attention to redress and reform hold relevance for our analysis of alternative approaches to dataset production. The dataset projects we present in this paper seek to repair content moderation by accounting for and learning from the experiences of those most impacted by online harms. It is an approach developed in response to the neglect of online harms and content moderation needs of global south languages, regions and communities. 

In responding to harms, interrupting neglect, and attempting to develop tools and datasets for redress, Tattle’s dataset projects offer rich grounds to draw on reparative justice as an analytical frame and pose the question: can data work be reparative? Bringing a reparative lens to examine the practices of data work invites the questions i) \textit{how}, if at all, might data work be reparative and ii) \textit{what} might a reparative data work seek to repair? In examining the reparative potential of data work through a case study of Tattle’s work, our interest does not lie in answering the question in the binary, definitively concluding whether or not data work can be reparative. Instead, we approach it as an open-ended question that offers an opportunity to consider how the goals and commitments of reparative justice can help inform and sharpen efforts to reimagine data work, dataset production and AI.

\section{Methods}


In this paper, we focus on Tattle’s work developing datasets in the context of oGBV, hate speech and sexual violence. As a civic tech organisation, Tattle works at the intersections of research and critical technical practice, and much of their work is highly collaborative, partnering with researchers and other civil society organisations. This paper draws on the lead author’s ethnographic engagement with Tattle from April to December 2024. As Tattle is a fully-remote organisation, ethnographic ‘fieldwork’ was primarily virtual, over Slack channels, online meetings and other collaborative digital tools. The lead author collaborated with Tattle to help develop guiding principles for compensation to dataset contributors and governance of the datasets. This involved 12 interviews with dataset contributors and Tattle team members, as well as ethnographic immersion, following and participating in Tattle’s safety datasets-related activities for eight months. 

In the section below, we describe the two dataset projects we examine --- a slurs lexicon dataset, and an LLM safety benchmark dataset. While our observations draw on these two specific dataset projects that the lead author engaged with during their fieldwork, the concerns we discuss build on and speak to Tattle’s broader experiences as a civic tech organisation attempting to bring a feminist orientation to their data and AI practice. 

\subsection{Following safety dataset projects at a civic tech organisation}

Tattle’s work on countering toxic online content began in 2021 with project <project name removed for peer review>, which addressed online gender-based violence (oGBV) on social media platforms, particularly in Indian languages where content moderation remains inadequate. They created an annotated Tweets dataset on gendered abuse and a lexicon of gendered slurs, by collaborating with women and queer researchers, practitioners, and activists with lived experience of gendered abuse. Their participation shaped both the conceptualisation and annotation of the datasets. This project demonstrated a feminist, participatory approach to AI dataset production, which forms the foundation for the two dataset projects discussed in this paper. 

The first project is the expansion of the slurs lexicon created under <project name removed for peer review>. There are very few resources \footnote{The List of Dirty, Naughty, Obscene, and Otherwise Bad Words (LDNOOBW) being one of the most widely used lists for filtering toxic language \cite{simonite_ai_nodate}} that document abusive slurs and phrases, and almost none for Indian languages, despite being essential for tools that filter out toxic language. Recognising this need, Tattle continued to maintain and expand the slur list as a separate dataset after the initial project concluded. At the time of our fieldwork, the slur list included more than 650 entries in Hindi, Tamil, Malayalam and Indian English with metadata such as level of severity, whether the slur is casually used, what makes it problematic, and qualify the slurs with categories such as gendered, sexualized, religion, sexual identity, ethnicity, political affiliation, caste, class, body shaming and ableist. We followed this project ethnographically, including the online contribution sessions held by Tattle in April 2024, with four feminist media organisations in India. These weekly, hour-long Zoom sessions involved 8–10 contributors collaboratively adding and annotating slurs in a shared Google Sheet. Small-group discussions at the start of each session helped build rapport and facilitate collaboration, feedback, and collective metadata creation. 

The second project is a benchmark dataset for LLM safety in Hindi, funded by ML Commons\footnote{ML Commons is an AI engineering consortium, involving stakeholders from across industry and academia. It aims to enable the advancement of AI, most notably through the development of benchmarks, by drawing on support, knowledge and resources from its stakeholders.} as part of its ambitious goal to build a state-of-the-art AI safety benchmark suite\footnote{A benchmark suite is a collection of standardized tests designed to evaluate the performance of computer systems or algorithms under specific conditions. It allows for consistent comparisons across different systems or models by using the same set of tasks and datasets.}. The benchmark (in version v1, released in late 2024) includes prompts to test LLM safety across 12 hazard categories\footnote{ The hazard categories are hate, defamation, non-violent crimes, privacy, suicide and self harm, sex-related crimes, violent crimes, intellectual property, child sexual exploitation, indiscriminate weapons, sexual content, and specialised advice \cite{ghosh_ailuminate_2025}}. Tattle was selected as a pilot project to create a dataset of 2000 prompts in Hindi for two hazards (hate and sex-related crimes). Through this project, Tattle aimed to demonstrate the value and indeed, need for a feminist orientation to safety benchmark, rather than relying on machine translations or a lay understanding of a language. 

Tattle brought together 23 contributors, including some past contributors, who use Hindi in their everyday lives and work across areas like social work, digital media, journalism, fact checking, psychology, and gender and feminist studies. Many belonged to or worked closely with marginalised communities in India and understood, very well, intersecting forms of harm related to gender, caste, religion, class, ethnicity, and sexuality. Tattle developed the prompts dataset through a series of six online workshops, each spanning 90 minutes. Tattle identified hate as the first hazard for the project; sex-related crimes (SRC) was collectively chosen as the second hazard. Dataset contributions (ie., prompt writing) were discussion-led, facilitated in small groups using shared documents, to build on each other’s prompts, share ideas, themes, topics and add variations. The lead author actively supported this project, helping Tattle develop the definitions and scope for the two hazards, analysing ML Commons’ safety framework, and refining participatory methods for prompt writing.

Across both projects, contributors documented abusive slurs, stereotypes, and actions targeting minoritised groups, enriching the datasets with critical nuance and contextual insight. Rather than analysing these contributions themselves, this paper focuses on Tattle’s approach to dataset production and the tensions it reveals.

\subsection{Data Analysis}
In this ethnographic study, the lead author was both an observer and active participant in the Hindi LLM safety project, contributing as a socio-technical researcher. Accordingly, this paper reflects an engaged positionality that helped shape the work under study. Our analysis of field notes, interviews, and related materials draws on interpretive traditions in socio-technical and ethnographic research \cite{dourish_implications_2006, singh_margins_2017, singh_whats_2025, taylor_gift_2003, tracey_intermediation_2024, vaghela_interrupting_2022, wang_please_2020}. We adopted an abductive approach \cite{timmermans_theory_2012} involving iterative reading and coding. In our preliminary analysis, we identified care, justice, and participation as recurring themes.Thinking with STS scholarship on feminist ethics, care, and repair, we refined the initial codes into broader themes of expansive data practices, recognition and value, and collective governance, discussed in Section 4. We shared an early draft of the paper with Tattle, and incorporated their reflections to add nuance to the tensions in their practice. We choose not to anonymise Tattle to ensure visibility and recognition for their work.

\section{A critical orientation to dataset production}

In both projects, the contributors were recognised as integral to building safety datasets rooted in the deeply contextual experiences of harms such as oGBV, sexual violence, and hate speech. The contributors who joined the two projects came from different professions ranging from social work, journalism, fact-checking, academic research, counselling, psychology, policy advocacy, and activism. Each contributor had direct lived experience of encountering harms or working closely with victims and survivors of harms such hate speech, oGBV, or sexual violence. It is these experiences that informed their contributions to the datasets. 

For each dataset project, Tattle identified and invited the contributors, through purposive sampling, based on the relevant languages and types of harms addressed in the dataset. The contributors came from different walks of life — some were in stable, full time employment, while others were freelancers or on short term contracts. further, they lived and worked in different parts of India, and came from diverse social backgrounds, including from marginalised caste, class, and gender communities. Tattle intentionally sought this diversity among its dataset contributors, to surface intersectional perspectives on the harms addressed in the datasets. Thus, the lived experience, social location, occupation and language proficiency of each contributor offered valuable and fairly unique but connected standpoints from which they considered and contributed data on hate speech, oGBV and sexual violence.

The role of the contributors, here, was not confined to providing ‘data’. Their input, drawing on contextual knowledge and on-ground experiences, shaped the datasets’ orientation to harms (such as oGBV and SRC) and enriched their approach to safety. It resulted in new labels, reframed metadata fields, and even redefining the scope of hazards. For instance, in the slurs lexicon project, contributors highlighted how certain slurs, like randi and rakhail among others, while being used as gendered and sexualised slurs, also demean sex work as a profession and suggested ‘profession’ as an additional category to label slurs\footnote{As an addition to existing categories: gendered, sexualized, religion, sexual identity, ethnicity, political affiliation, caste, class, body shaming and ableist.}. In the safety benchmark project, contributors similarly questioned the dataset’s scoping of “sex-related crimes” as a hazard and advocated re-orientating the hazard’s focus from sex-related \textit{crimes} to sexual \textit{violence}. Such engagement with contributors, beyond the confines of annotation tasks and data, is both rare and a significant departure from the prevailing modes of dataset production, particularly in content moderation contexts. 

To engage deeply on topics like hate speech and sexual violence, bring with it, its own set of harms. Given the nature of the content in the datasets, such as slur words, derogatory phrases, and stereotypes, they posed mental health and well-being risks. To address these risks, Tattle aimed to limit the duration of exposure to harmful content. Contributors had access to the dataset documents only during the online workshops. Similarly, workshop sessions were spaced out and limited to once a week, for not more than 90 minutes. Further, Tattle partnered with a mental health service provider in the safety benchmark project, to provide access to trauma-informed counselling and therapy to all its contributors. 

Supporting the engaged participation of a diverse set of contributors also required attending to their varying levels of digital proficiency and access. In this regard, Tattle adopts an intentionally open-ended and flexible approach to data collection and annotation. Tattle sought to lower barriers to participation for contributors, nurture discussions and collaboration among them, and be attentive to the toll of working with toxic content. Dialogue and discussions were integral to this approach. For instance, in the safety benchmark project, prompt writing was discussion-led and happened in groups. Some groups were heavily discussion-oriented with contributors actively discussing reported cases of hate crime or sexual violence, stereotypes and biases faced by specific identity groups, etc. These discussions then formed the basis for prompt writing by other contributors and the Tattle team in subsequent sessions. 

A discussion-centric approach thus allowed for contributors to participate and provide valuable input to the dataset, even if it did not always take the form of data. This was especially important for contributors who were not comfortable accessing and working on shared online documents. It also created a collaborative environment where people were not engaging with toxic content alone, by themselves. Creating space for discussions and collaborative contributions were intentional choices that Tattle made, responding to feedback from past contributors, on the process and experience of participating in dataset activities. These efforts are rooted in care and recognition of the varied levels of digital access and language proficiency among contributors and were integral to being attentive to the harms arising from the nature of the data work involved in the projects.

This approach, taking a feminist orientation to dataset production, illustrates the possibilities to develop deeply contextual datasets by engaging with a diverse range of contributors and learning from their situated knowledges. More crucially, perhaps, for how we understand the role of the data worker or contributor, it demonstrates the possibilities for broadening the scope of human involvement in dataset production for AI. By troubling the norms of what counts as data, annotation or contribution to enable a larger, more integral role for dataset contributors, Tattle’s approach reorients data work as a site for repair and redress. Grounded in feminist commitments, this reorientation must attend not only to producing contextual datasets on safety, but also to critically examining conditions, and relations of power under which the datasets are built. In this section, we elaborate on the challenges confronted by Tattle as they sought to address these questions. Specifically, we shed light on challenges in advancing just reward and compensation for data work, and collective dataset governance and why these are less recognised, yet crucial, avenues for repair in data work. 

\subsection{Recognising and valuing contributions}

Recognising the relevance of lived experience for safety datasets and taking an expansive view of data to engage with it, in turn, raises questions for Tattle about how the input and labours of those contributing to the datasets might be valued and rewarded? In this section, we discuss Tattle’s efforts to confront and address this in their practice.  

This has not been an easy question to answer for Tattle. First, the compensation practices in the data enrichment industry remain opaque, with research and media reportage showing data work pay in the global south often falling below minimum wage. Advocating for a living wage therefore becomes the focal point for research and advocacy groups campaigning for fair conditions of work \cite[p.~102]{graham_fairwork_2020}. As a result, there is little guidance available for projects or organisations like Tattle that acknowledge the issue of exploitative wages and are committed to going beyond minimum wage or living wage. This is especially pertinent for their work as they seek lived experience as a source of valuable and expert input for datasets. If the dataset contributors were regarded as experts in a sense, it was important to also reflect that in the compensation they received.  

Second, the variety in the nature of datasets, tasks, and contributor profiles complicate attempts to develop a standardised framework for compensation. For instance, in one of the earliest safety datasets that Tattle produced, the contribution task involved annotating about 8000 Tweets to detect abusive language for which compensation was on a per-task basis. This was intended to avoid placing any rigid demands or limits on the time contributors took for a task. In contrast, the safety benchmarking project in 2024 involved prompt writing that was structured as a series of six workshops for which compensation was on the basis of time (ie., the number of workshops that contributors joined) rather than by number of prompts written. As a project with an exploratory agenda, it prioritised more open-ended, free-flowing discussions over reaching a targeted number of prompts. In our interviews with Tattle’s dataset contributors, they did not mind or have a strong preference for either mode of compensation. For many, the motivation to contribute was the chance to inform and intervene in the problem space of online safety that the datasets tackled. To them, compensation was secondary, as noted below by one of the contributors we interviewed: 

\begin{quote}
    I see this [my contribution] as something I do for both myself and society… we are all a part of society… something like this, I see it as a form of collective effort… so it is not just about the payment…. But another thing: I have finished a Doctorate… I work in an engineering college as an assistant professor. In some sense, I am not depending on that money. I recognise this is a form of privilege… I have what I need, so this is more like an extra for me, like a gift or a prize. But I acknowledge that not everyone may view it like that. My financial status may allow for it. 
\end{quote}

However, not everyone took the same view. In the interviews with contributors, most of them expressed satisfaction about the compensation they had received, motivated primarily by the project’s goals rather than the monetary reward. However, a past contributor, an author and tech policy consultant, had declined to join the safety benchmark project\footnote{ In the case of the slur list expansion project, the contributors who participated were affiliated with one of 4 feminist media organisations that Tattle partnered with, for the project. Here, Tattle paid the organisations for their partnership and contribution to the dataset, alongside other collaborative work.} due to higher expectations of compensation. The contributors joining these projects came from different walks of life, living and working under different material conditions, in different sectors and across different parts of India. The group of contributors in the safety benchmark project included academics, researchers, social workers, journalists, fact checkers, writers, and psychologists. All contributors were compensated on the basis of an equal rate of 1200 INR (approximately 13.5 USD) per hour (which is approximately about 10 times the industry compensation rate in India) or 1800 INR (approximately 20 USD) for each workshop of 90 minutes. But, not all of them were in full-time employment or held permanent positions. This variation in qualifications, geographic location, sectoral norms, and material conditions add to the challenge of computing fair compensation. It is a tension that not only Tattle faces but also social enterprises providing data work services \cite[p.~42]{murgia_code_2024}. At the end of the safety benchmark project, there was money leftover from the contributor compensation budget. This was in part because it was hard to know, at the start, exactly how many contributors would join and for how many workshops. The hourly compensation rate was set at 1200 INR to accommodate more contributors than there were eventually. At the end of the project, Tattle distributed the remaining funds as additional compensation of 932 INR ((approximately 10.5 USD)) per workshop to the contributors. 

Further, any efforts to improve compensation for data work also involved convincing funding bodies. Tattle’s efforts to value contributions above the minimum wage and exploitative market rates has often meant that their proposal or bid for dataset projects can be seen as expensive. This was the case, as they learnt, for the safety benchmark project. Even as their proposal was selected, it was as a pilot study. It was evident to Tattle that they were less likely to be funded for taking this approach to build larger-scale datasets. It would be considered too expensive. This was also highlighted in our interview with a senior data worker, with close to a decade of data work experience and having founded her own data work firm recently. She observed that new companies starting out in the sector quote very low rates for data work projects in an attempt to outbid competitors. Well-intentioned actors who may quote higher prices to compensate data work fairly, ultimately lose out in an industry driven by a race to the bottom. Tattle had no interest in being yet-another data sourcing company. Their interest, and indeed commitment, in taking on dataset projects lay in examining and demonstrating the possibilities for (and benefits to be gained from) taking a feminist approach to safety datasets.  

Our efforts to conceptualise a framework for compensation surfaced more questions than answers. But these questions hold wider relevance beyond Tattle’s work. Amidst growing efforts to draw on annotator subjectivity \cite{jiang_understanding_2021, patton_annotating_2019, rottger_two_2022, santy_nlpositionality_2023, sap_annotators_2022} and diversity \cite{aroyo_dices_2023, kapania_hunt_2023, meer_annotator-centric_2024}, how their role and contributions to datasets are valued is a crucial question. In the Fairwork project’s 2025 ratings for cloudwork\footnote{ Cloudwork platforms here refer to online platforms like Appen, Amazon MTurk, and Prolific that enlist data workers to perform short-term, on-demand tasks.} platforms \cite{fairwork_fairwork_2025}, only 4 out of the 16 platforms evaluated paid workers at least the local minimum wage, revealing that the road to just reward for data labours remains long. While efforts to enforce better compliance of labour laws are crucial to guarantee minimum standards in compensation, they are only the first step towards fair pay. Recognising human input to datasets and the associated labours as crucial and valuable requires rewarding it as such. Tattle’s ongoing efforts to confront this demonstrate that dataset projects seeking high quality, expert input, or contextual knowledge cannot sidestep the implications they bear on the recognition and value of data work.  

\subsection{Rethinking the contributors’ relationship to the datasets} 

Besides considering how contributions are rewarded, reimagining recognition, from a feminist orientation, also brings into question on what terms contributors participate in the datasets. This invites a new set of questions such as whose views and voices shape dataset decisions such as its licensing, plans for their expansion or maintenance, what rights contributors have over their contributions, whether they may revoke or refuse the use of their contributions in specific instances or towards specific ends, the scope for individual rights and collective decision-making in dataset production, and what the role of Tattle might be as custodians of the datasets. In this section, we present Tattle’s experiences in grappling with some of these questions and highlight the need for suitable models of governance and stewardship that address the tensions arising from feminist modes of engagement in dataset production. 

Tattle’s engagement with this line of enquiry stems from having to consider future plans for the datasets they produced. First, online toxic content evolves constantly featuring new slurs, coded language, and references to contemporary events that respond and adapt to the prevailing discourse at a given time. This means that for datasets like the slur list to remain relevant for training or benchmarking language models, they require to be regularly updated. This brings into focus how and who maintains the datasets, who contributes, and what these roles entail. Second, Tattle also began to explore how the slurs dataset could be made available to social media and content platforms. Though India represents a major market and a large user base for social media platforms, they rarely have in-house expertise to understand and address online abuse in the Indian context, making the slurs dataset a highly relevant and crucial resource for content moderation and safety on these platforms. 

Engaging with platform companies entails considering on what terms the project might do so. At a time of indiscriminate scraping and appropriation of online content by AI models, should a dataset, developed as a feminist endeavour, be licensed for commercial use? Should it be made freely available in the interest of wider adoption and easy access? Or should it resist being available freely to companies that have long ignored their needs and experiences? If the dataset is monetised, what distinguishes Tattle and their contributors from vendors conducting outsourced data work for requesters? If platforms were to pay, how might that earnings be used? Should it, for instance, be shared amongst dataset contributors like royalty pay in recognition of their contributions? What other rights and responsibilities, if at all, should contributors have in the governance of the datasets? 

These are important questions about the relationship between the dataset, its contributors and tech companies that represent potential adopters or users. They are also real decisions that Tattle grapples with in considering the future of the datasets they produce. Lowering barriers to participation and facilitating deeper engagement with contributors go a long way in enriching the datasets and helping gather more situated input. However, without confronting the implications for the governance of the datasets, such efforts, while thoughtful, risk remaining a largely transactional mode of operation. This is why it was (and remains) important for Tattle to reflexively examine how the datasets they produce might be governed, though it proved challenging to operationalise, as we show below.  
\subsection{Challenges collective dataset governance}
First, the contributors involved in Tattle’s dataset projects change over time, complicating possibilities for collective decision-making. Tattle developed the <project name removed for peer review> dataset in 2022. The slurs dataset grew out of the <project name removed for peer review> work and has been maintained and growing since 2023, through occasional contribution activities and sprints. In September 2024, Tattle undertook the related but separate project to develop the prompt dataset for AI safety benchmarking. Over the course of 3 years, dataset contribution activities took place sporadically, at irregular intervals and lasting a few weeks each time, primarily contingent on Tattle securing funding to support contributors. Further, shifts in the immediate focus of the dataset like working on specific languages or specific themes like sex-related crimes brings in new contributors as well. This meant the pool of contributors neither remains stable nor is constantly engaged with the datasets. Given the bursts of activity within which contributors get involved, their continued availability or indeed, interest in being involved in matters of dataset governance is not a given. Their sporadic involvement poses a challenge for collective decision-making and governance that is typically premised on sustained participation. 

Second, while the contributors were undoubtedly important to the dataset projects, it was not always the case that the projects were important to them. Not all contributors considered AI safety or online safety as a priority. Many of them were engaged in the frontlines of social work, activism, and journalism, and did not automatically view online harms and safety as an urgent problem to be addressed. In some cases, as the research lead at Tattle pointed out in an early interview and quoted below, there were other larger and more imminent threats they faced: 

\begin{quote}
    “In the scheme of things that they worry about, online abuse is not a very big concern. For example, some of the journalists have dealt with physical violence. Online violence to them, it's not the biggest problem! One of them very frankly asked, why is this even a problem that we’re trying to solve! While the online abuse they face is a lot more than what a lot of other people might face, they face a lot of different kinds of hardships as well that are more important [to them].“
\end{quote}

It was Tattle that recognised the relevance of their work, experiences and knowledge to the dataset projects and invited them to participate as contributors. The contributors viewed the issue of online abuse as an extension of systemic issues such as domestic violence, caste atrocity, violence against the queer community and other kinds of harms they encountered and addressed in their work and activism. The stakes of their participation were primarily rooted in responding to these issues. Concerns of platform accountability or appropriation of knowledge commons did not automatically feature as a priority. Through interviews with the contributors, we learnt that most contributors viewed choices concerning engagement with platforms as outside their realm of knowledge or expertise. They suggested that it was for Tattle (or “the people leading the project”\footnote{Some contributors related the dataset work to the people they were in touch with, rather than Tattle as an organisation. In our interviews, they either used names of lead contacts or vague references like “they”, “the project team”, etc.}) to make those decisions. To the limited extent that they shared their thoughts on this, they erred on the side of making in-roads with platforms like Instagram, X and YouTube than holding those platforms to account, even if it meant that the fruits of their labour would be freely available to the powerful platforms that have neglected their problems. As one of them noted during the interview: 

\begin{quote}
    “If having them get access to this [dataset] for free and sort of incorporating it in there like moderation, frameworks, etc ensures or even marginally benefits women, queer people, trans people, or those from diverse realities that we want to support through this effort, I don't think there's a tension there at all. I mean, of course, it's unfair, but like that is the reality that we've always lived in. So in that sense, not engaging with big tech platforms is not an option. They're not gonna do it on their own because if they wanted to do it on their own, they would have done it by now.” 
\end{quote}

While most contributors echoed similar, even if sometimes reluctant, sentiments of taking the datasets where they might be most needed (ie., social media platforms), one of them sought to question that direction of travel. This came from a critical scholar of technology who was previously involved with the <project name removed for peer review> dataset project as a researcher. When she returned as a dataset contributor to the LLM safety benchmark project, she raised critical questions about the licensing terms of the dataset and advocated for more independence and control over dataset governance. This critical stance over the governance of datasets and engagement with powerful tech industry actors, while valued by Tattle, was an exception amongst contributors.  

Reimagining dataset governance, just as with strengthening recognition for dataset contributors, remains an open question for Tattle. As an initiative building on limited resources and contributions from a shifting community of participants with varied motivations and expertise, the grounds for sustained participation or collective decision-making cannot be taken for granted. Yet, it remains a relevant and important pursuit.

\section{Discussion: An agenda for repair in data work}

Tattle represents a highly unusual site to examine data work. They are not in the business of providing data enrichment services; nor are they invested in building AI products driven by commercial, profit-making interests. Their dataset projects can be situated at the sparsely occupied intersections of research and critical technical practice, primarily focussed on responding to oGBV and online harms. Yet, their work constitutes a relevant case study of data work, providing valuable insights and pertinent questions for interrupting prevailing extractive and exploitative tendencies. Our study surfaces productive tensions and generative possibilities that remain less recognised in critical data work scholarship, and yet crucial to confront. Specifically, the challenges concerning just reward for data workers and collective dataset governance directly point towards concrete avenues for redress in data work. 

Having elaborated on these tensions in some detail above, we now return to the question we outlined at the start of the paper: can data work be reparative? Drawing on the engagement with Tattle, we reflect on what it can tell us about the reparative potential of data work and AI. Specifically, we consider how data work might be reparative and what reparative data work might strive to repair. In examining this we draw attention to the work that lies ahead for resetting the relations of accountability in AI and their implications for the future of data work and AI practice. Our analysis aims to offer a different, justice-oriented view of accountability in data work and AI, pointing to avenues for change, rather than offering prescriptive policy recommendations for AI accountability. Thinking with the analytical frame of reparative justice sharpens the focus on defining shared standards and forms of responsibility to interrupt prevailing norms of dataset production and to place at the centre, not AI or datasets, but those most harmed by them. 

\subsection{Resetting the relations of accountability in AI}

Reparations are concerned with not only how harms and violence are redressed but also what is being repaired in the process. It does not seek to annul harms but to affirm a new baseline that repairs the relationships at the root of the harms, one that recognises the standing of those most impacted by the harms to demand accountability, and the obligation of those who inflicted them to acknowledge, accept responsibility, and make amends\cite{walker_what_2010}. It is a commitment to establish shared standards for accountability that can form the basis for new relationships of trust, respect, and reconciliation. For Walker, it is this resetting of the relations of accountability that lies at the core of reparative justice and what it aims to repair \cite{walker_making_2015}. We draw on this reparative attention to relations of accountability as a useful, critical orientation to consider what reparative approaches to AI might aim to repair. 

The datasets produced by valuing situated knowledge and engaging with it through data work may reduce the possibilities of future harms but it does not repair them. Without the authority to hold the platforms accountable, the inclusion of those harmed in the datasets alone does not make any amends. It is not the gaps and oversight in datasets that a reparative approach must strive to repair but the lack of accountability that precisely leads to those harms in the first place. Making amends then lies in establishing shared standards that allow those most affected by AI to hold AI companies accountable. 

Such reparative visions of accountability, however, are a far cry from the status quo of the tech sector. Anthropologist Diana Forsythe’s seminal research on the construction of knowledge \cite{forsythe1993engineering} and construction of work \cite{forsythe1993construction} in AI draws attention to the hierarchies of expertise and labour within the production of AI, that lead to “narrow and brittle” technologies. Her analysis highlights the striking absence of recognition of technology production as a relational process, requiring accountable ties between different forms of knowledge and labour involved in the process. These hierarchies and imbalances continue to characterise contemporary computing, including in online safety efforts and AI production. Meta’s Trusted Partner programme, for instance, was meant to serve as a collaboration with “expert organisations that represent the voices and experiences of marginalised users around the globe” to strengthen online safety for at-risk users \cite{meta_bringing_2023}. However, a review conducted by some of the Trusted Partners reveal an under-resourced initiative that relies on its partners taking on the significant burden and risk of reporting issues without a commitment to redress or accountability from Meta \cite{internews_safety_2023}. Indeed, such asymmetric, extractive relations are not even registered or scrutinised by  formal accountability mechanisms such as audits (cite broken bus). As Cobbe et al \cite{cobbe2023understanding} note, accountability measures fail to adequately account for political economic dynamics that underpin AI systems, and thus largely remain fragmented and dislocated in AI supply chains \cite{widder_dislocated_2023}. 

The resulting neglect of harms and tokenistic attempts to address them signal the need for a reparative rethinking of accountability. It is in this context that Tattle’s exploration of governance models for its datasets becomes crucial. In Tattle’s specific case, it entails considering how the fruits of a feminist endeavour should and can be made available to AI and platform companies, especially those that have neglected concerns of safety so far. We argue that the surfaced challenges concerning dataset governance must be read, not merely as procedural or practical constraints but as productive tensions in resetting the relations of accountability. This resonates with justice-oriented calls to examine the role of collectivity in data governance \cite{dencik2025collectivity}. As both a practice and a field of academic inquiry, data governance is rooted in individualist frames that prioritise procedural safeguards while failing to account for how contemporary data systems actively reconfigure social relations to consolidate power. In response, Dencik et al \cite{dencik2025collectivity} advocate a data-justice led approach to governance that explicitly centres the relational and collective dimensions of data and AI systems. Our case  study directly responds to this call, demonstrating the centrality of collective governance in fundamentally reorienting safety dataset production.

\subsection{Implications for dataset and AI practice}

While our entry point to engaging with Tattle’s dataset projects was a focus on the labor and work practices shaping dataset production, the insights from this research hold broader relevance beyond data work. In this section, we outline the implications our findings and analysis hold for efforts in other related spheres, in particular for efforts focused on multilingual LLMs, data and knowledge commons, and labour justice in AI. Here, by implications, we do not refer to technical or policy proposals that can be readily adopted in other contexts; rather we identify critical links between our research and other ongoing struggles against the deepening power imbalances and lack of accountability in the trajectories of AI proliferation. By highlighting these links, we indicate new research directions from these interconnected challenges. In that sense, the implications we outline below contribute towards an agenda building for alternative futures of data and AI practice.   

The recent proliferation of LLMs has fueled an interest in building multi-lingual systems and extending LLM capabilities for languages that have been historically underrepresented in digital spheres. However, with the majority of such efforts originating from actors, institutions in the global north and fueled by resources from large corporations, they are rife with tensions of extraction, appropriation and acute power asymmetries \cite{smart2024socially, bird2024must}. In response to these tensions, communities in global south regions have sought to assert their language sovereignty and epistemic authority by developing new licensing models for their language datasets. Licenses like the Esethu License \cite{rajab2025esethu}, the Nwulite Obodo Open Data License (NOODL) \cite{okorie_licensing_2024, okorie_its_2025} and the Kaitiakitanga License \cite{hao_new_2022, te_hiku_media_tehikumediakaitiakitanga-license_nodate} have been specifically crafted to deter extractive practices, and prioritise reciprocity, sovereignty and community stewardship. Such efforts resonate with Tattle’s endeavour presented above, and highlight the role of collective governance in advancing epistemic justice and agency. 

Licensing for community-led data initiatives is, however, witnessing a shift with the proliferation of AI. In open source and open knowledge communities, content and data licenses like Creative Commons and ODbL were devised about two decades ago to enable reuse and nurture the digital commons. However, the “techno-legal openness” they afford has proven inadequate in curbing the appropriation of commons resources to train proprietary AI models that do not contribute any value back to the commons and only benefit the already powerful companies that develop and control these models \cite{chandrasekhar_legal_2025}. The questions and tensions surfaced in the context of Tattle, then, are pertinent not only for safety datasets and minoritised languages, but also for threats facing the future of data and knowledge commons.

Besides licensing, establishing the relations of accountability in AI has also taken other forms. The 2023 Hollywood strikes led by actors and screenwriters unions SAG-AFTRA\footnote{Sag-Aftra or the Screen Actors Guild–American Federation of Television and Radio Artists is a labour union representing over 160,000 actors, artists and media professionals.} and WGA\footnote{WGA or the Writers Guild of America represents over 11000 writers across two labour unions based in the East and West Coast of the US.} against the use of AI to undermine, mimic or replace their creative labour mobilised worker power to govern the use of AI in the creative industry \cite{grohmann_worker-led_nodate}. More recently, in March 2026, a class action lawsuit was brought against Grammarly, a writing assistance tool for reviewing grammar and spelling, for misappropriating names, and identities of hundreds of authors, writers, journalists, and editors in their recent AI-driven feature called ‘Expert Review’ that presented editing suggestions as if they were from established, well-known authors \cite{wiredGrammarlyFacing}. While these union and legal actions against AI practices might not seem immediately relevant to the context of safety datasets examined in this paper, we highlight critical ways in which they begin to intersect with each other. These cases begin to show how, in the context of AI, concerns about labour are deeply entangled with concerns about data misappropriation, consent, and extraction. The research we present in this paper makes this entanglement explicit, and highlights how epistemic justice and labour justice intersect in reimagining data and AI futures, and why they cannot be viewed in isolation.

These licenses, protests, demands, and refusals constitute concrete articulations of relations of accountability within the contexts in which they intervene. They strive to repair the specific ways in which their contributions and concerns are undermined or neglected, signifying, as Walker notes, \textit{“the possibility of a relationship of accountability and reciprocity where there has been none”} \cite[p.~218 emphasis in the original text]{walker_making_2015}. In responding to past harms and modelling a better future, the central work of reparations lies in the present, here and now, in igniting hope that shared standards of respect are possible to achieve \cite{walker_what_2010} and that our worlds can be (re)made to flourish on the basis of just, caring relations \cite{papadopoulos_ordinary_2023}.

\section{Limitations}

While our research affords in-depth insights into the structures, practices and tensions that underpin this work, our focus is limited to a single case study, of one organisation’s efforts to do data work differently. Here, we recognise that Tattle is neither alone in attempting to reorient data work, nor is their approach the only way to do so. Other initiatives are examining ways to strengthen the possibilities for a democratic, agential work environment \cite{disalvo_when_2024} and worker mobilising in data work \cite{miceli_methodological_2025} as pathways to reimagine data work. Our research adds to this nascent but important line of enquiry through the case study of Tattle. Similarly, we highlight the importance of rewarding data work fairly and enabling models of collective governance of datasets. However, these are not the only avenues of repair for data work. Researchers and worker unions engaged with data work in different parts of the world have highlighted various issues concerning working conditions, wages, and mental well-being that are crucial to repair \cite{yang_guilds_2024, zhang_making_2025, fairwork_fairwork_2025, posada_deeply_2024, sambasivan_everyone_2021, wang_whose_2022}. The agenda we build for a reparative data work intersects with these (and other) labour concerns at different points of the AI supply chains. It complements and strengthens the demands and goals of other allied struggles towards alternative data and AI futures.

\section{Conclusion}

The growing body of literature on data work has predominantly focused on its work practices and labour concerns. This is undoubtedly important and urgently required. We extend this body of work, in this paper, by inviting attention to a second but related set of themes concerning the epistemic tensions and the politics of knowledge production in data work. The case study of Tattle’s dataset projects offers an enriching opportunity to examine these themes in depth, especially highlighting the tensions that arise in translating alternative visions for AI into actual practice. Bringing a reparative lens to this analysis offers a productive push to confront crucial questions including whose knowledge is important and relevant to datasets, how it is sought and valued, on what terms and set by whom. Thinking with reparations not only surfaces these questions but also demands that any attempts to answer them must prioritise direct relief, and strengthening accountability to those most harmed by the neglect, oversight and exclusion animated in the current modes of dataset production. Reparative justice offers a bold vision for responsibility that may seem impossible, but is precisely why, as the history of reparations shows us, it remains crucial to advancing repair and justice.

\section{Generative AI Usage Statement}

The manuscript was prepared without the use of generative AI tools. The contents of the submission, including its structure and style, is solely the work of the authors, and did not involve the use of LLMs or other generative AI tools.

\section{Acknowledgments}

Thank you to Tattle Civic Tech and the dataset contributors for being a part of this research and sharing their ongoing work, motivations, experiences, and struggles. Contributions to this chapter by Alex Taylor were made in his capacity as a BRAID Research Fellow and funded by the BRAID programme and the UK’s public funder, AHRC / UKRI (award AH/X007146/1).

\bibliographystyle{ACM-Reference-Format}
\bibliography{ref}

\end{document}